\documentclass[twocolumn, journal]{IEEEtran} 
\makeatletter 
\def\ps@headings{%
\def\@oddhead{\mbox{}\scriptsize\rightmark \hfil \thepage}%
\def\@evenhead{\scriptsize\thepage \hfil \leftmark\mbox{}}%
\def\@oddfoot{}%
\def\@evenfoot{}} 
\makeatother 
\usepackage{latexsym,times,epsf,amsmath,amssymb,amsfonts,graphicx}
\usepackage{epstopdf}
\usepackage{graphicx}
\usepackage{subfigure}
\usepackage{multirow}
\usepackage{cite}
\usepackage{setspace}
\usepackage{comment}
\usepackage{float}

\usepackage{algpseudocode}
\usepackage{algorithm}
\usepackage{mathrsfs}
\usepackage{array}
\renewcommand{\baselinestretch}{0.94}

\begin{document}
\title{Satellite Based Positioning Signal Acquisition at Higher Order Cycle Frequency}
\author{Pengda Wong}
\maketitle

\begin{abstract}
The acquisition of the signal from the satellite based positioning systems, such as GPS, Galileo, and Compass, encounters challenges in the urban streets, indoor. For improving the acquisition performance, the data accumulation is usually performed to improve the signal-to-noise ratio which is defined on the second order statistics. Different from the conventional approaches, the acquisition based on higher order cyclostatistics is proposed. Using the cyclostatistics, the estimation of the initial phase and Doppler shift of the signal is presented respectively. Afterwards, a joint estimator is introduced. The analysis in this paper is performed on GPS signal. Indeed, the proposed estimation method can be straightforwardly extended to acquire the signal from the other satellite positioning systems. The simulation and experiment results demonstrate that the proposed signal acquisition scheme achieves the detection probability of 0.9 at the CNR 28dBHz.
\end{abstract}

\section{Introduction}
\label{sec:introduction}

With the extension satellites based positioning and navigation service, the GPS receiver encounters the problem of the signal acquisition in the challenging environments, such as indoor, on urban street and in woods with dense leaves~\cite{Tsui2001}. In these environments, the GPS signal becomes weak and can hardly be acquired by the receivers. To solve the problem, the studies on the weak GPS signal acquisition methods are widely performed. Generally, these GPS signal acquisition methods can be divided into two main categories. First, piling up GPS data is adopted to increase SNR~\cite{Psiaki01,Elders-Boll2004,Yu2007,Huang2011a,Yiming2011}. Second, mitigating the interference from the noise, the jamming signal and the unexpected GPS signals~\cite{Wang2011,Morton2007,Huang2011} is another approach to improving GPS signal acquisition performance.

Coherent accumulation is a classical method of acquiring weak GPS signal acquisition. Even though the SNR increases with the extending accumulated GPS signal length, the coherently accumulated signal length does not goes beyond 10ms due to the data bit transition. To improve the SNR furthermore, Psiaki~\cite{Psiaki01} studied two non-coherent GPS signal accumulation methods, ‘full-bit’ and ‘half-bit’  GPS signal acquisition methods, which noncoherently pile up the GPS data to improve the GPS signal acquisition performance. In the condition that the carrier-to-noise (CNR) is larger than zero, the GPS signal noncoherent accumulation will generate the higher SNR. However, the accumulated GPS data duration can not be extended without limits. The noncoherent GPS signal accumulation will induce square loss; and the Doppler shift will also limit the accumulated GPS signal length extension. Thus, the GPS signal acquisition performance improvement by the noncoherent GPS signal accumulation is limited. At the direction of increasing SNR by piling up the GPS data, differential coherent accumulation method is also employed~\cite{Elders-Boll2004,Yu2007}. The differential coherent accumulation based acquisition method sums up the products of two adjacent coherent results. The GPS signal acquisition method increases the Doppler shift tolerance which is also insensitive to the data bit transition. Thus for differential coherent accumulation the constraint on the GPS signal data length extension is not so strict as the coherent accumulation which length is shorter than 10ms in usual cases. However, the shortage of the differential coherent accumulation is obvious that the SNR gain efficiency is not high enough compared with coherent accumulation method in the condition of the small Doppler shift.

Wang etc.~\cite{Wang2011} employed the noise subspace tracking algorithm to reduce the degradation from the interference which is a kind of spatial filtering method. The noise subspace tracking method depends on an antenna array to implement beam forming to mitigate the interference. Inevitably, the antenna array will increase the cost of the GPS receiver. Morton etc.~\cite{Morton2007} studied the GPS signal self-interference mitigation method by subspace projection. However, in the study of Morton, the non-orthogonality between the different GPS signal is neglected of, that is, the cross-correlation result between different pseudorandom codes is not exact zero. Huang and Pi~\cite{Huang2011} studied the interference between the different GPS signals which is called the near-far effect in GPS signal acquisition. Three different kinds of solutions to the near-far problem are proposed. Unfortunately, there is no one general method which can handle all the three kinds of the near-far problems. Besides that, the coexistence of multiplicative and additive noise in GPS signal is deeply studied~\cite{Huang2013b}.

Huang etc.~\cite{Huang2009} proposed a GPS signal detection algorithm which employed Duffing chaotic oscillator to detect the weak GPS signals. The GPS signal detection algorithm utilizes the immunity to noise and sensitivity to the periodical signals. GPS signal has two periods possessed by the carrier and the pseudorandom code. Meanwhile, the noise is of no any periodicity in usual cases. The external periodical force from the periodic input signal will change the state of the chaotic oscillator by which the GPS signal is detected. However, there is a gap between the algorithm study and the hardware implementation since the chaotic oscillator based GPS signal acquisition method demands for the quite heavy computation burden. Also, Liu etc.~\cite{Liu2009} and Sahmoudi etc.~\cite{Sahmoudi2009} studied the degraded GPS signal tracking algorithm in the challenging environments with dense multipath and jamming signal. As we know, the tracking algorithm in the challenging environments makes sense only after the weak GPS signal is successfully acquired. The studies on improving hardware of GPS receiver are performed in~\cite{Huang2009a,Huang2012,Huang2010a}. The mobile states assisted positioning performance improvement is studied in~\cite{Huang2014}.

Indeed, the GPS signal possesses cyclostationary feature which enables the GPS receiver distinguish the expected GPS signal from the background noise and jamming signal. To the best knowledge of the author, few publications refers to utilizing the cyclostastionary feature to acquire the weak GPS signal.

In the 1950s, cyclostationary feature was proposed to characterize the statistics which is non-stationary but periodic. Since the 1980s Gardner W.A. has performed wide and deep research work on the cyclostatistics and the related applications~\cite{Gardner1975,Gardner1986,Gardner1988,Gardner1988a,Gardner1991,Gardner1992,Gardner1993}. Until now, there are huge amount of signal processing algorithms based on cyclostatistics in the various fields. However, as the author can refer, there is no publication on GPS signal detection based on the cyclostationary feature.

In this paper, the cyclostationary feature of the GPS signal is analyzed firstly; after that, the GPS signal detection method based on cyclostatistics is proposed and the detection performance is analyzed; also the initial pseudorandom code phase and the Doppler shift estimation methods are studied; in the end part of this article the simulations and experiments are carried out to test the effectiveness of the proposed GPS signal acquisition scheme.

\section{Cyclostationary Feature and Cyclic Spectrum of GPS Signal Equations}
\label{sec: cyclic feature}

The GPS signal is represented by $s(t)$. The $s(t)$ is the time period of 1\textit{ms} is given by 
\begin{equation}
g(t)=\sum_{k=0}^{1022}c(k)q(t-kT_c-t_0)\cos(2\pi f_0t+\theta_0).
\label{eq: def of s(t)}
\end{equation}
where $T_c$ denotes the time duration of one pseudorandom code chip. $q(t)$ is the square wave with duration, $q(t)=1$ for $0\leq t\leq T_c$, otherwise, $q(t)=0$. $t_0$ is the initial pseudorandom code phase. $f_0$ is the signal carrier frequency. $\theta_0$ is the initial phase of the carrier.



To investigate the cyclostationary feature of GPS signal, we calculate the mean and the variance of $s(t)$

In one period of C/A code, due to ergodic feature, the mean value of $s(t)$ is approximately calculated by
\begin{equation}
\begin{aligned}
M_{s}&=\int_0^{T_0}s(t)dt\\
&=\int_0^{T_0}\sum_{k=0}^{1022}c(k)q(t-kT_c-t_0)\cos(2\pi f_0t+\theta_0)dt\\
&\approx 0,
\end{aligned}
\label{eq: mean of s(t)}
\end{equation}
where $T_0$ denotes the period of C/A code.

The autocorrelation of the GPS signal $s(t)$ is written as 
\begin{equation}
\begin{aligned}
R_{s}(t,\tau)&=\int_t^{t+\Delta t}s(t)s(t+\tau)dt\\
\end{aligned}
\label{eq: autocor 1 of s(t)}
\end{equation}

Since $s(t)$ is periodic at the period of $T_0$, that is, $s(t+T_0)=s(t)$, we have
\begin{equation}
\begin{aligned}
R_{s}(t+T_0,\tau)_{\Delta t}&=\int_t^{t+\Delta t}s(t+T_0)s(t+T_0+\tau)dt\\
&=\int_t^{t+\Delta t}s(t)s(t+\tau)dt\\
&=R_{s}(t,\tau)_{\Delta t}
\end{aligned}
\label{eq: autocor 1 of s(t)}
\end{equation}

Based the results in the previous several steps, the mean value of GPS signal is constant and autocorrelation function is periodic versus time. Therefore, GPS signal is cyclostationary. The derivation is according to the definition by Gardner. Because of the periodicity in the GPS signal autocorrelation function, Fourier series of the periodic function can be calculated as follows,
\begin{equation}
\begin{aligned}
R_{s}(t,\tau)&=\sum_{n=-\infty}^{+\infty}R_G^{\frac{n}{T_0}}(\tau)e^{j2\pi\frac{n}{T_0}t}
\end{aligned}
\label{eq: Fourier decomp of Rs}
\end{equation}
where $\frac{n}{T_0}$ is the cyclic frequency, $n\in\mathbb{Z}$ and 
\begin{equation}
\begin{aligned}
R_G^{\frac{n}{T_0}}(\tau)=\int_{\Delta t}R_G(t,\tau)e^{-j2\pi\frac{n}{T_0}t}dt
\end{aligned}
\label{eq: Fourier component}
\end{equation}

For derivation convenience, let $\alpha$ denote the cyclic frequency, $\alpha=\frac{n}{T_0}$. $R_G^{\frac{n}{T_0}}(\tau)$ is thecyclic-autocorrelation function at cyclic frequency at $\alpha$. According to the definition, the cyclic spectrum, denoted by $S_G^{\alpha}(f)$, is calculated as follows,
\begin{equation}
\begin{aligned}
S_G^{\alpha}(f)=\int R_G(t,\tau)e^{-j2\pi f\tau}dt
\end{aligned}
\label{eq: cyclic spectrum cal}
\end{equation}

Until now, the background of cyclostatistics and the cyclostationary feature of GPS signal are introduced. Next, we will present the cyclostatistics based initial code phase estimation method.

\section{Cyclic-spectrum Based Initial Code Phase Estimation}
\label{sec: initial phase est.}

To successfully acquire GPS signal, we need to obtain the two information, initial pseudorandom code phase and the Doppler shift. As the prime acquisition taks, initial phase estimation based on cyclicspectrum is introduced in this section.

\subsection{Conventional Initial PN Code Phase Estimation}
\label{subsec: conventional est.}

Remember $g(t)$ denotes the ideal GPS signal. The received one is denoted by $r(t)$. Let $\tau$ denote the pseudorandom~(PN) phase difference between $g(t)$ and $r(t)$ and $\tau<|T_0|$. With the definitions, the received signal $r(t)$ is written as follows
\begin{equation}
\begin{aligned}
r(t)=A\cdot g(t-D)+\xi(t)
\end{aligned}
\label{eq: r(t) def}
\end{equation}

The conventional initial phase estimation is based on the peak detection on the correlation between the ideal signal $g(t)$ and the received one $r(t)$. The correlation can be straightforwardly obtained as follows,
\begin{equation}
\begin{aligned}
R_{R,G}(\tau)=A\cdot R_{G}(\tau-D)+R_{\Xi,G}(t,\tau).
\end{aligned}
\label{eq: cor r and g}
\end{equation}

To implement the peak detection, we smoothly change the value of $\tau$ within the duration of $T_0$. Since the noise $\xi$ is uncorrelated with the signal $g(t)$, $R_{\Xi,G}$ approaches to zero in the high SNR case. Thus, the correlation $R_{R,G}$ is determined by the autocorrelation function $R_{G}$. When $\tau=D$, $R_{G}$ reaches the maximum value, so $R_{R,G}$ does. The peak detection is completed.

The conventional acquisition method can easily implemented. However, when  GPS encounters severely degraded noise, $R_{\Xi,G}$ no longer approaches to zero; thus the autocorrelation result will submerge in the background noise and the peak detection can not be successfully achieved.

\subsection{Initial Phase Estimation Method Based on Cyclic-spectrum Correlation}
\label{subsec: cyclicspec based phase est}

As shown previously, GPS signal is cyclostationary, while the noise does not. In the interference channel, jamming signal might be cyclostationary or not. Even for cyclostationary jamming signal, its cyclostationarity is different from that of the expected GPS signal. Therefore, we are able to detect the GPS signal from the background with strong noise and interference.  Concretely speaking, the cyclic-spectrum is utilized to estimate the initial PN code phase.


 Let $g^{\prime}(t)$ denote the replica of $g(t)$ with a phase shift $\delta$, that is, $g^{\prime}(t)=g(t-\delta)$. $S_{G^{\prime},G}^{\alpha}(f)$ denotes the cyclic-spectrum calculated from $g(t)$ and $g^{\prime}(t)$, and $S_{R,G}^{\alpha}(f)$ denotes the cyclic-spectrum calculated from the received signal $r(t)$ and $g(t)$. The cyclic-spectrum $S_{R,G}^{\alpha}$ is calculated by
\begin{equation}
\begin{aligned}
&S_{R,G}^{\alpha}(f)=\mathcal{F}\left\{R_{R,G}^{\alpha}(\tau)\right\}\\
&=\mathcal{F}\left\{\lim_{T\rightarrow\infty}\frac{1}{T}\int_{-T}^T \left(\begin{matrix}(g(t-D+\frac{\tau}{2})+\xi(t))\\ \cdot g^*(t-\frac{\tau}{2})\end{matrix}\right)e^{-j2\pi\alpha t}dt\right\}\\
&\overset{(a)}{=}\mathcal{F}\left\{\lim_{T\rightarrow\infty}\frac{1}{T}\int_{-T}^T \left(\begin{matrix}g(t-\frac{D}{2}+\left(\frac{\tau}{2}-\frac{\tau}{2}\right))\\ \cdot g^*(t-\frac{D}{2}-\left(\frac{\tau}{2}-\frac{\tau}{2}\right))\end{matrix}\right)e^{-j2\pi\alpha t}dt\right\}\\
&{=}\mathcal{F}\left\{R^{\alpha}_G(\tau-D)e^{-j\pi\alpha D}\right\}\\
&{=}S^{\alpha}_G(f)e^{-j2\pi\left(\frac{\alpha}{2}+f\right) D}\\
\end{aligned}
\label{eq: S r,g}
\end{equation}
where $(a)$ follows that the cyclostatistic of noise is zero when $\alpha\neq0$.

Similarly, we calculate $S_{G^{\prime},G}^{\alpha}$ which is listed as follows,
\begin{equation}
\begin{aligned}
&S_{G^{\prime},G}^{\alpha}(f)=S^{\alpha}_G(f)e^{-j2\pi\left(\frac{\alpha}{2}+f\right) \delta}\\
\end{aligned}
\label{eq: S g prime,g}
\end{equation}

The similarity between the two cyclic-spectrum is calculated to complete the estimation of the initial PN code phase. Let $D_{ec}$ denote the similarity which is calculated as follows,
\begin{equation}
\begin{aligned}
D_{ec}&=\int_{-\infty}^{\infty}{S_{G^{\prime},G}^{\alpha}(f)}^*S_{R,G}^{\alpha}(f)df\\
&=\int_{-\infty}^{\infty}|S_{G}^{\alpha}(f)|^2e^{-j2\pi\left(\frac{\alpha}{2}+f\right)\left(\delta-D\right)}df\\
\end{aligned}
\label{eq: Dec}
\end{equation}

From (\ref{eq: Dec}), when
\begin{equation}
-j2\pi\left(\frac{\alpha}{2}+f\right)\left(\delta-D\right)\rightarrow 0,
\end{equation}
$D_{ec}$ approaches the maximum. According to the result, the initial pseudorandom code phase delay $D$ can be obtained by calculating the maximum absolute value of the cyclic-spectrum correlation.

Until now, the theoretic proof on the initial phase delay estimation based on the cyclic-spectrum is completed. However, the calculation performed in the derivation is far from a practical case. Next, we will propose a practical scheme to implement the cyclic-spectrum based initial phase estimation.

\subsection{Practical Scheme of the Cyclic-spectrum Based Initial Phase Estimation}
\label{subsec: practical cal}

According to the definition, the cyclostatistics $R_{R,G}^{\alpha}(\tau)$ is calculated by
\begin{equation}
\begin{aligned}
&R_{R,G}^{\alpha}(\tau)=\lim_{T\rightarrow\infty}\frac{1}{T}\int_{-\frac{T}{2}}^{\frac{T}{2}}\left(\begin{matrix}(r(t+\frac{\tau}{2})e^{-j\pi\alpha\left(t+\frac{\tau}{2}\right)})\\ \cdot(g(t-\frac{\tau}{2})e^{-j\pi\alpha\left(t-\frac{\tau}{2}\right)})\end{matrix}\right)dt\\
\end{aligned}
\label{eq: R r,g alpha cal}
\end{equation}

According to the previous analysis results, $R_{R,G}^{\alpha}(\tau)$ is periodic at the period of $T_0$. The cyclic-spectrum $S_{R,G}^{\alpha}(f)$ is calculated as follows,
\begin{equation}
\begin{aligned}
&S_{R,G}^{\alpha}(f)=\frac{1}{T_0}\int_{T_0}R_{R,G}^{\alpha}(\tau)e^{-j2\pi\tau f}d\tau\\
&=\frac{1}{T_0}\int_{T_0}\frac{1}{T}\lim_{T\rightarrow\infty}\int_{t-\frac{T}{2}}^{t+\frac{T}{2}}\left(\begin{matrix}(r(\zeta+\frac{\tau}{2})e^{-j\pi\alpha\left(\zeta+\frac{\tau}{2}\right)})\\\cdot(g(\zeta-\frac{\tau}{2})e^{-j\pi\alpha\left(\zeta-\frac{\tau}{2}\right)})
\end{matrix}\right)  d\zeta e^{-j2\pi\tau f} d\tau\\
&=\lim_{T\rightarrow\infty}\frac{1}{T}\int_{t-\frac{T}{2}}^{t+\frac{T}{2}}\frac{1}{T_0}\int_{T_0}\left(\begin{matrix}
(r(\zeta+\frac{\tau}{2})e^{-j\pi\alpha\left(\zeta+\frac{\tau}{2}\right)})\\ \cdot(g(\zeta-\frac{\tau}{2})e^{-j\pi\alpha\left(\zeta-\frac{\tau}{2}\right)})
\end{matrix} \right) d\tau e^{-j2\pi\tau f}  d\zeta\\
&=\lim_{T\rightarrow\infty}\frac{1}{T}\int_{t-\frac{T}{2}}^{t+\frac{T}{2}}\frac{1}{T_0}\int_{T_0}\left(\begin{matrix}\left(r(\zeta+\frac{\tau}{2})e^{-j\pi\left(\frac{\alpha}{2}+f\right)\left(\zeta+\frac{\tau}{2}\right)}\right)\\ \cdot\left(g(\zeta-\frac{\tau}{2})e^{-j\pi\left(-\frac{\alpha}{2}+f\right)\left(\zeta-\frac{\tau}{2}\right)}\right)^*
\end{matrix}\right)d\tau  d\zeta\\
&=\lim_{T\rightarrow\infty}\frac{1}{T}\int_{-\infty}^{\infty}\frac{1}{T_0}\int_{T_0}\left( \begin{matrix}
(r(\zeta+\frac{\tau}{2})e^{-j\pi\left(\frac{\alpha}{2}+f\right)\left(\zeta+\frac{\tau}{2}\right)})\\ \cdot(g(\zeta-\frac{\tau}{2})e^{-j\pi\left(-\frac{\alpha}{2}+f\right)\left(\zeta-\frac{\tau}{2}\right)})^*
\end{matrix}\right) d\tau  d\zeta\\
&~~~~~\cdot\otimes rec_T(t)
\end{aligned}
\label{eq: S r,g alpha cal2}
\end{equation}
where
\begin{equation}
\begin{aligned}
rec_T(a)=\left\{\begin{matrix}\frac{1}{T}, & |a|\leq\frac{T}{2}\\
0, & |a|>\frac{T}{2}\end{matrix}\right.
\end{aligned}
\label{eq: rect def}
\end{equation}

For derivation simplicity, we define two auxiliary variables, $R_{T_0}(t,f)$ and $G_{T_0}(t,f)$ as follows,
\begin{equation}
\begin{aligned}
&R_{T_0}(t,f)=\frac{1}{\sqrt{T_0}}\int_{T_0}r(t+\frac{\tau}{2})e^{-j2\pi(\frac{\alpha}{2}+f)(t+\frac{\tau}{2})}d\tau\otimes rec_T(t)\\
&=R_{T_0}^{\prime}(t,f+\frac{\alpha}{2}),
\end{aligned}
\label{eq: R t0}
\end{equation}
and
\begin{equation}
\begin{aligned}
&G_{T_0}(t,f)=\frac{1}{\sqrt{T_0}}\int_{T_0}r(t+\frac{\tau}{2})e^{-j2\pi(\frac{-\alpha}{2}+f)(t-\frac{\tau}{2})}d\tau\otimes rec_T(t)\\
&=G_{T_0}^{\prime}(t,f-\frac{\alpha}{2}),
\end{aligned}
\label{eq: G t0}
\end{equation}
where
\begin{equation}
\begin{aligned}
&R_{T_0}^{\prime}(t,f)=\frac{1}{\sqrt{T_0}}\int_t^{t+T_0}r(\zeta)e^{-j2\pi f\zeta}d\zeta\\
&G_{T_0}^{\prime}(t,f)=\frac{1}{\sqrt{T_0}}\int_t^{t+T_0}g(\zeta)e^{-j2\pi f\zeta}d\zeta\\
\end{aligned}
\end{equation}

Substituting (\ref{eq: R t0}) and (\ref{eq: G t0}) into (\ref{eq: S r,g alpha cal2}), we have the cyclic-spectrum as follows,
\begin{equation}
\begin{aligned}
S_{R,G}^{\alpha}(f)&=\lim_{T\rightarrow\infty}\frac{1}{T}\int_{-\frac{T}{2}}^{\frac{T}{2}}R_{T_0}(t,f){G_{T_0}(t,f)}^*d t\otimes rec_T(t)\\
&=\lim_{T\rightarrow\infty}\frac{1}{T}\int_{-\frac{T}{2}}^{\frac{T}{2}}R_{T_0}^{\prime}(t,f){G_{T_0}^{\prime}(t,f)}^*d t
\end{aligned}
\label{eq: S r,g 3}
\end{equation}

Now, the practical scheme in analog time domain is presented. For the application in the widely used digital circuits, we next transform the analog practical scheme into discrete time form. In the discrete time domain, the cyclic-spectrum is calculated as follows,
\begin{equation}
\begin{aligned}
&S_{R,GD}^{\alpha}[m]=\lim_{N\rightarrow\infty}\frac{1}{N}\sum_{n=1}^{N}R_{T_0D}[n,m]{G_{T_0D}[n,m]}^*\\
=&\lim_{N\rightarrow\infty}\frac{1}{N}\sum_{n=1}^{N}R_{T_0D}^{\prime}\left[n,m+\frac{\Phi}{2}\right]{G_{T_0D}^{\prime}\left[n,m-\frac{\Phi}{2}\right]}^*\\
=&\lim_{N\rightarrow\infty}\frac{1}{N}\sum_{n=1}^{N}\left(\begin{matrix}
\left(\frac{1}{\sqrt{M}}\sum_{m=n+1}^{n+M}r[n]e^{-j2\pi (m+\frac{\Phi}{2})pt_s}\right)\\ \cdot{\left(\frac{1}{\sqrt{M}}\sum_{m=n+1}^{n+M}g[n]e^{-j2\pi (m-\frac{\Phi}{2})pt_s}\right)}^*
\end{matrix} \right) 
\end{aligned}
\label{eq: S r,g discrete}
\end{equation}
where $f_s$ is the sampling frequency and $t_s$  is the sampling period; $\Phi$ denotes the shift in the frequency domain caused by cyclic frequency $\alpha$ which is calculated by $\Phi=2\cdot floor\left(\frac{\alpha}{2f_s}\right)$.

\begin{figure*}[h]
\vspace{0 mm}
 \centering
  \includegraphics[width=1\textwidth]{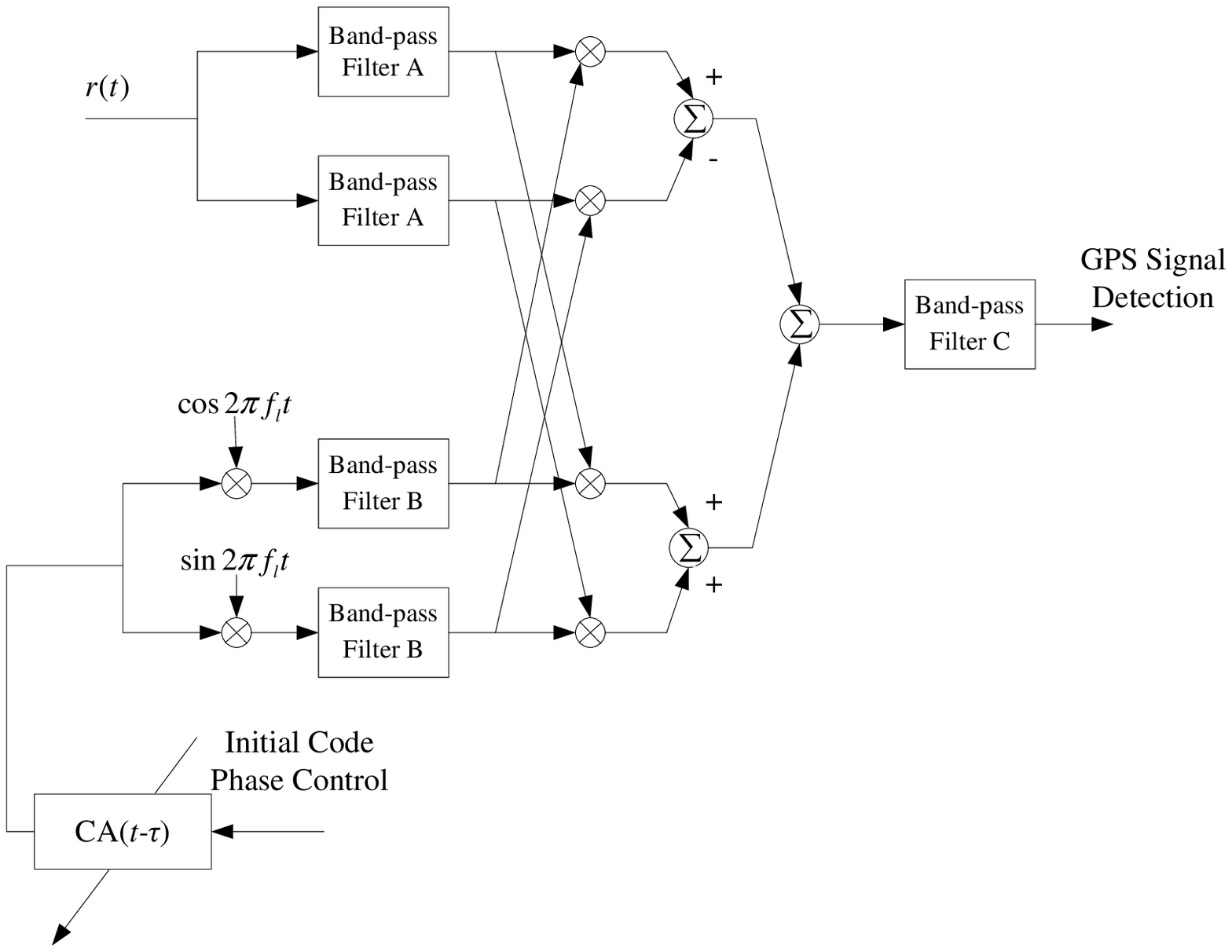}
 \caption{GPS signal acquisition block diagram based on cyclic-spectrum correlation}
\label{fig: cyclic-spec acquisition}
\vspace{0 mm}
 \end{figure*}

For a more intuitive understanding on the practical scheme, we present a block diagram of the practical cyclic-spectrum based initial phase estimation scheme. The block diagram is shown in Fig.~\ref{fig: cyclic-spec acquisition}

In Fig.~\ref{fig: cyclic-spec acquisition}, the central frequency of the band-pass filter A is located at $\frac{\alpha}{2}+f$ with bandwidth of $\Delta f$; the central frequency of the band-pass filter B is located at $-\frac{\alpha}{2}+f$ with bandwidth $\Delta f$; the central frequency of band-pass filter C is located at $\alpha$ with bandwidth of $\frac{1}{\Delta t}$.

In processing shown in Fig.~\ref{fig: cyclic-spec acquisition}, to alleviate the frequency leaking problem, frequency smoothing is performed. The bandwidth of the frequency smoothing window is equal to $\Delta f$. We define $I$ as, 
\begin{equation}
\begin{aligned}
I=\left\lfloor\frac{\Delta f}{f_s}\right\rfloor.
\end{aligned}
\label{eq: I def}
\end{equation}

Let $\tilde{S}_{R,GD}^{\alpha}[m]$ denote the cyclic-spectrum after the smoothing operation which is calculated as follows,
\begin{equation}
\begin{aligned}
&\tilde{S}_{R,GD}^{\alpha}[m]
=\frac{1}{N}\sum_{n=1}^{N}\frac{1}{I}\sum_{i=-\frac{I-1}{2}}^{\frac{I-1}{2}}\left(\begin{matrix}
(R_{T_0D}^{\prime}\left[n,m+\frac{\Phi}{2}+i\right])\\ \cdot({G_{T_0D}^{\prime}\left[n,m-\frac{\Phi}{2}+i\right]}^*)
\end{matrix} \right) \\
\end{aligned}
\label{eq: S r,g discrete smooth}
\end{equation}

\begin{figure}[h!]
     \begin{center}
       \subfigure[{GPS signal cyclic-spectrum before smoothing}]{%
            \label{fig:before smooth}
          \includegraphics[width=1.6in]{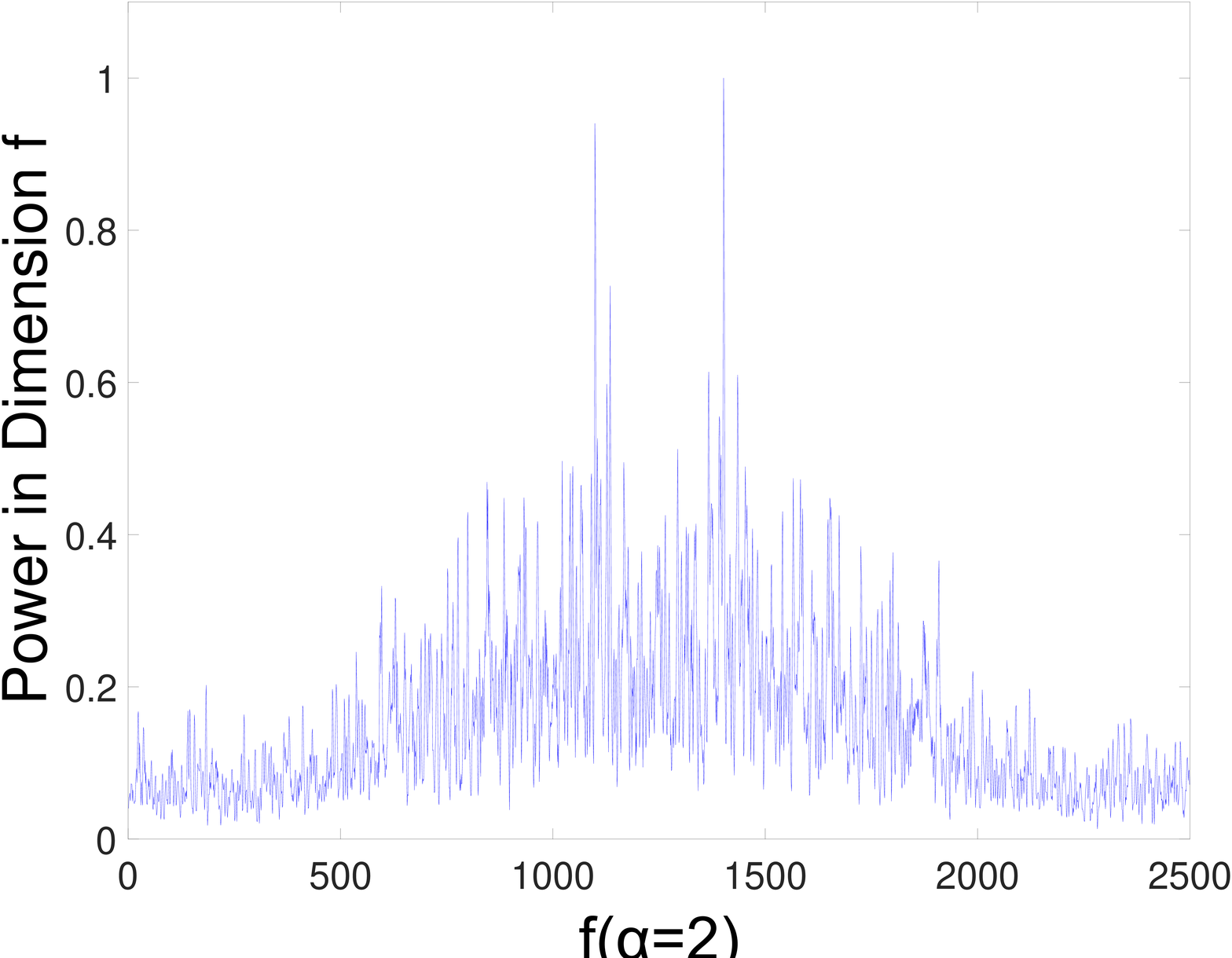}
    }
     \subfigure[GPS signal cyclic-spectrum after smoothing]{%
        \label{fig:after smooth}
       \includegraphics[width=1.6in]{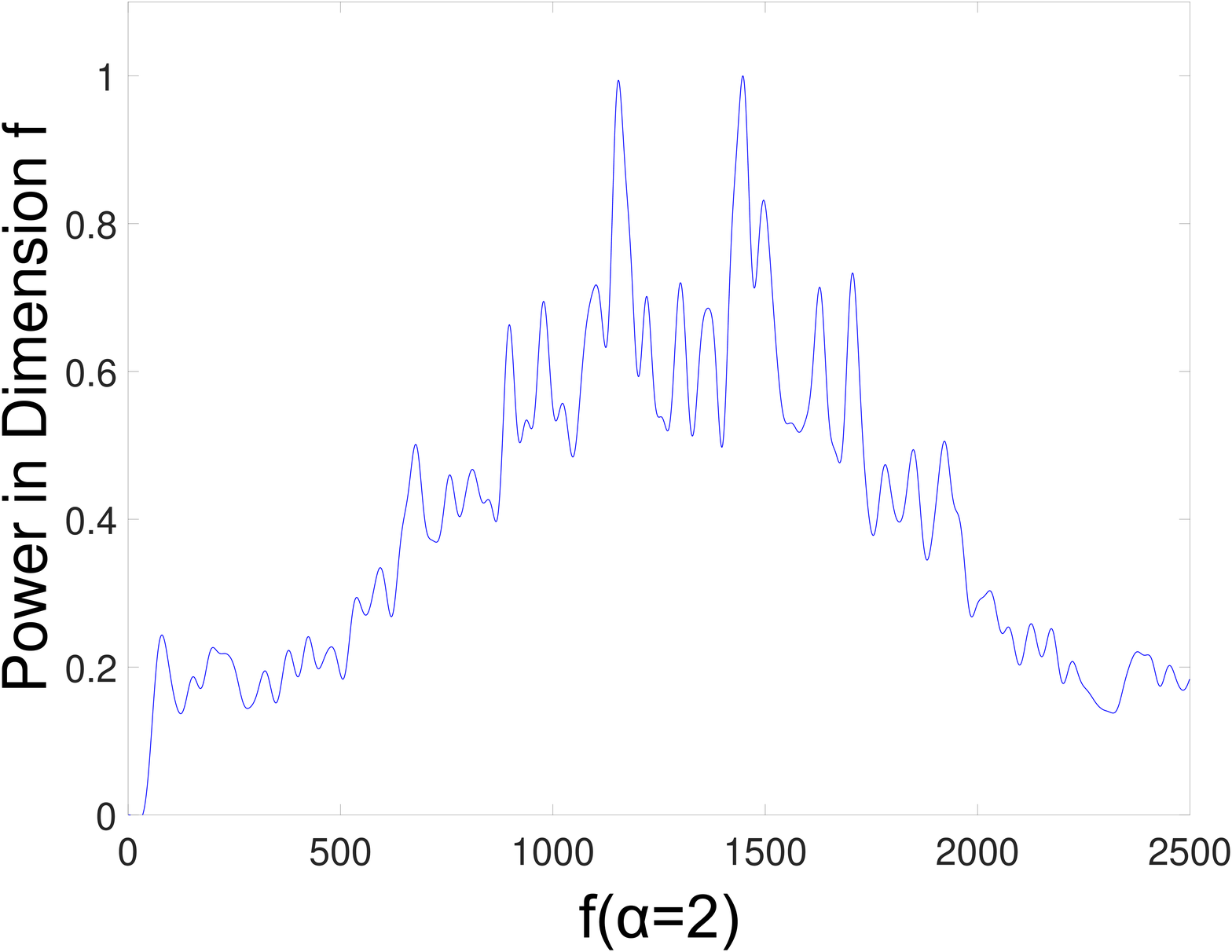}
    }\\ 
\end{center}
\caption{GPS signal cyclic-spectrum smoothing demo}
\label{fig:smoothing}
 \end{figure}

Fig.~\ref{fig:before smooth} and Fig.~\ref{fig:after smooth} illustrate the cyclic-spectrum of the GPS signal before and after the frequency smoothing. From Fig.~\ref{fig:smoothing}, we can observe the frequency leaking problem is released after the smoothing operation.

\subsection{Initial PN Code Phase Iterative Estimation Scheme}
\label{subsec: iterative est}

The block processing based initial phase estimation method is presented in the previous subsection. The block operation requires the large duration of data which generates processing delay. To solve the problem, we propose a iterative structure based initial phase estimation scheme in this subsection.

Remember that the received GPS signal in the discrete time domain is written as $r[n]=g[n-D]+\xi[n]$. According to the conclusion in~\cite{So1994a}, $g[n-D]$ can be approximately expanded by a set of \textit{sinc} functions as follows,
\begin{equation}
\begin{aligned}
g[n-D]=\sum_{i=-P}^P sinc(i-D)g[n-i],
\end{aligned}
\label{eq: g D decomp}
\end{equation}
where $sinc(x)$ is equal to $\frac{\sin(x)}{x}$ for $x\neq0$ and 1 for $x=0$; $P$ is a positive integer larger than $D$.

We calculate the cyclostatistics from $r[n]$ and $g[n]$ as follows,
\begin{equation}
\begin{aligned}
R_{R,G}^{\alpha}[n_{\tau}]=\left<\left(g[n-D]+\xi[n]\right)g[n-n_{\tau}]e^{-j2\pi\alpha n t_s}\right>_{\Delta t},
\end{aligned}
\label{eq: R r,g iter}
\end{equation}
where $\left<\cdot\right>_{\Delta t}=\frac{1}{\Delta t}\int_{\Delta t}$; $\Delta t$ is equal to the data length in the calculation, $\Delta t=N\cdot t_s$; and $n=\lfloor\frac{\tau}{t_s}\rfloor$.

The noise $\xi$ is not cyclostationary for $\alpha\geq1$. Thus, we have $R_{\Xi,G}^{\alpha}=0$ and $R_{R,G}^{\alpha}$ is rewritten as follows,
\begin{equation}
\begin{aligned}
R_{R,G}^{\alpha}[n_{\tau}]=\left<g[n-D]g[n-n_{\tau}]e^{-j2\pi\alpha n  t_s}\right>_{\Delta t},
\end{aligned}
\label{eq: R r,g iter2}
\end{equation}

Substituting (\ref{eq: g D decomp}) into (\ref{eq: R r,g iter2}), we have
\begin{equation}
\begin{aligned}
&R_{R,G}^{\alpha}[n_{\tau}]=\left<\begin{matrix}\\ \left(\sum_{i=-P}^P sinc(i-D)g[n+\frac{n_{\tau}}{2}-i]\right)\\ \cdot e^{-j\pi\alpha n  t_s}g[n-\frac{n_{\tau}}{2}]e^{-j\pi\alpha n  t_s}\end{matrix}\right>_{\Delta t}\\
&=\sum_{i=-P}^P \left(\begin{matrix}
sinc(i-D)e^{-j\pi\alpha i  t_s}\\ \cdot\left<g[n+\frac{n_{\tau}}{2}-i]e^{-j\pi\alpha (n-i)  t_s}g[n-\frac{n_{\tau}}{2}]e^{-j\pi\alpha n  t_s}\right>_{\Delta t}
\end{matrix} \right)  \\
&=\sum_{i=-P}^P sinc(i-D)e^{-j\pi\alpha i  t_s}R_{G}^{\alpha}[n_{\tau}-i],
\end{aligned}
\label{eq: R r,g iter3}
\end{equation}

Next, we define the cost function built on the cyclostatistics as follows,
\begin{equation}
\begin{aligned}
\varepsilon(k,n_{\tau})=&R_{R,G}^{\alpha}[n_{\tau}]\\
-&\sum_{i=-P}^P sinc(i-\hat{D}[k])e^{-j\pi\alpha i  t_s}R_{G}^{\alpha}[n_{\tau}-i],
\end{aligned}
\label{eq: cost func def}
\end{equation}
where $\hat{D}[k]$ denotes the estimation of the phase delay $D$ after $k$-th iterative computation.

From (\ref{eq: cost func def}), we have the square of the cost function as follows,
\begin{equation}
\begin{aligned}
J=\sum_{n_{\tau}=-N_{\tau}}^{N_{\tau}}|\varepsilon(k,n_{\tau})|^2,
\end{aligned}
\label{eq: cost func sq}
\end{equation}

Next, we calculate the derivate function of $J$ versus the estimated initial PN code phase delay $\hat{D}[k]$,
\begin{equation}
\begin{aligned}
&\hat{\nabla}[k] =\frac{\partial J}{\partial\hat{D}[k]}\\
&=2\sum_{n_{\tau}=-N_{\tau}}^{N_{\tau}}\varepsilon(k,n_{\tau})\sum_{i=-P}^P f \left(i-\hat{D}[k]\right)R_{G}^{\alpha}[n_{\tau}-i],
\end{aligned}
\label{eq: cost derivative}
\end{equation}
where $f(x)=\frac{\cos(\pi x)-sinc(x)}{x}$ and $N_{\tau}$ is the maximum initial phase shift.

Afterwards, we have the iterative code phase estimation method which is under the rule of minimum square error. The method is shown below,
\begin{equation}
\begin{aligned}
\hat{D}[k+1]=\hat{D}[k]-\mu\hat{\nabla}[k],
\end{aligned}
\label{eq: iter eq}
\end{equation}
where $\mu$ is the step in the iterative computation.

\begin{figure*}[h]
\vspace{0 mm}
 \centering
  \includegraphics[width=1\textwidth]{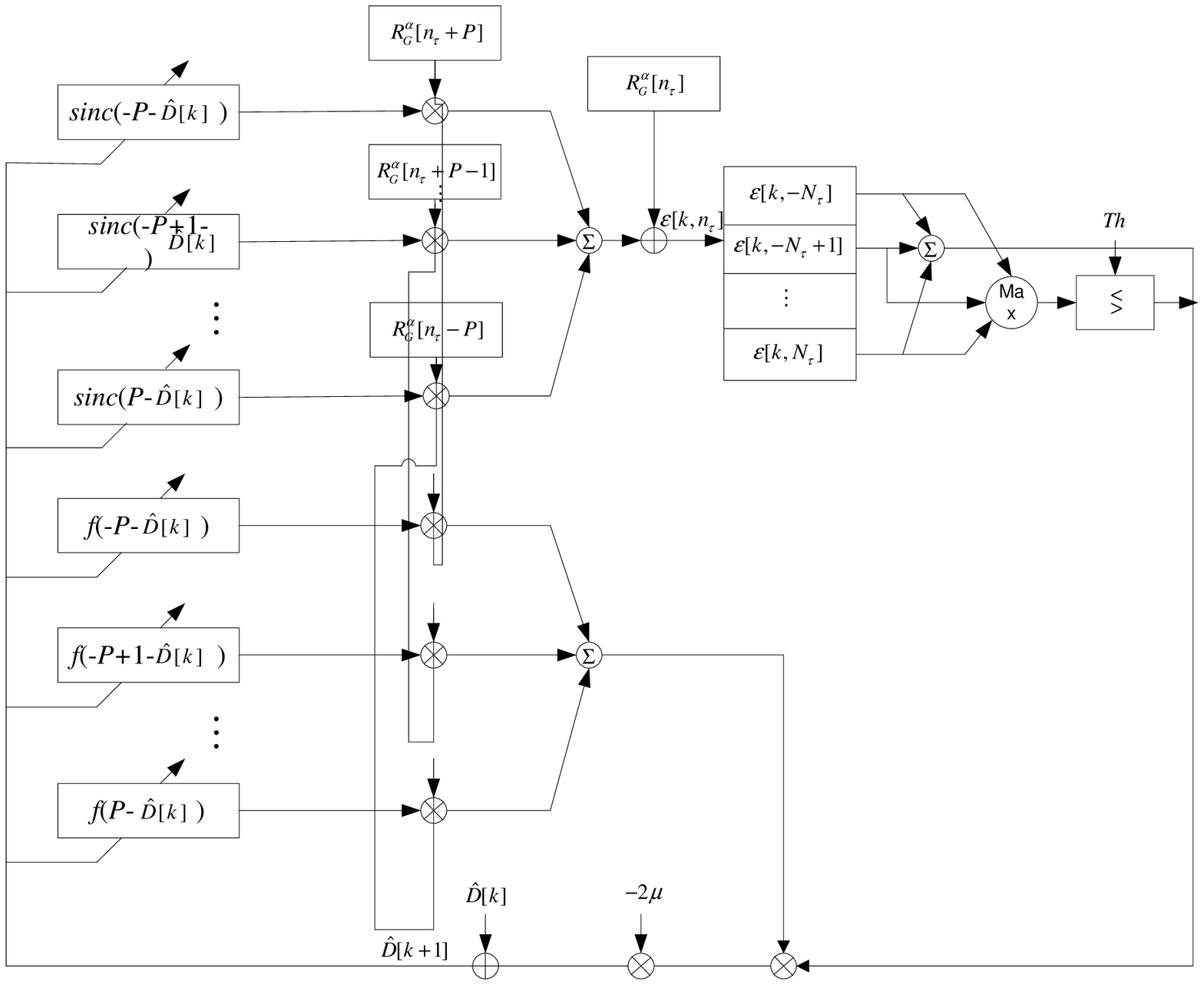}
 \caption{Iterative initial phase estimation block diagram}
\label{fig: iter est bd}
\vspace{0 mm}
 \end{figure*}

To present a intuitive impression, Fig.~\ref{fig: iter est bd} illustrates the block diagram of the iterative estimator.

\section{Doppler Shift Estimation Based on Cyclic-spectrum}
\label{sec: doppler est.}

For analysis simplicity, we first consider the GPS signal with Doppler shift, but no phase delay. From such an ideal signal, we estimate the Doppler using a cyclic-spectrum based method. In the next section, we will extend the method to acquire the practical GPS signal which is with both Doppler shift and phase delay.  

Still, let $r(t)$ denote the GPS signal which is written as follows,
\begin{equation}
\begin{aligned}
r(t) = g(t)e^{j2\pi f_d t}+\xi(t),
\end{aligned}
\label{eq: gps Doppler only}
\end{equation}
where $f_d$ is equal to the Doppler shift value.

\subsection{Conventional Doppler Estimation Method}
\label{subsec: conventional doppler}

The conventional Doppler shift estimation is based on measuring the similarity between $r(t)$ signal $l(t)$ with respect to the second order statistics. With the consideration of the periodicity of PN code, the correlation function of GPS signal can be written as 
\begin{equation}
\begin{aligned}
R(t,\tau) = \frac{1}{T}\int_{t}^{t+T}\left[g(\zeta)e^{j2\pi f_d\zeta}+\xi{\zeta}\right]g(\zeta-\tau)e^{-j2\pi f_d(\zeta-\tau)}d\zeta,
\end{aligned}
\label{eq: gps autocor}
\end{equation}
where $T$ is $n$ times the PN code period, $T=n\times T_0$.

Since the GPS signal $g$ is uncorrelated with the noise $\xi$ and the mean of $\xi$ is zero, 
\begin{equation}
\begin{aligned}
R(t,\tau) = \frac{1}{T}\int_{t}^{t+T}g(\zeta)g(\zeta-\tau)e^{j2\pi (f_d-f_{dl})\zeta+j2\pi f_{dl}\tau}d\zeta,
\end{aligned}
\label{eq: gps autocor2}
\end{equation}

Let  $\Delta f$ denote the frequency difference between the Doppler shift $f_d$ and $f_{dl}$, $\Delta f=f_d-f_{dl}$. $R(t,\tau)$ is further derived as follows,
\begin{equation}
\begin{aligned}
&R(t,\tau) = \frac{1}{T}e^{j2\pi f_{dl}\tau}\int_{t}^{t+T}g(\zeta)g(\zeta-\tau)e^{j2\pi \Delta f\zeta}d\zeta\\
=& \frac{1}{T}e^{j2\pi f_{dl}\tau}\int_{-\infty}^{\infty}\left(\begin{matrix}
g(\zeta)g(\zeta-\tau)e^{j2\pi \Delta f\zeta}\\
\cdot(u(\zeta-t)-u(\zeta-t-T))
\end{matrix} \right) d\zeta\\
\overset{(a)}{=}& \frac{e^{j2\pi f_{dl}}\tau}{T}\int_{-\infty}^{\infty}\left(\begin{matrix}
g(\zeta)g(\zeta-\tau)e^{j2\pi \Delta f\zeta}\\
\cdot (u(\zeta-t+\frac{T}{2})-u(\zeta-t-T-\frac{T}{2}))
\end{matrix}\right) d\zeta\\
\end{aligned}
\label{eq: gps autocor3}
\end{equation}
where $(a)$ follows the periodicity. 

Furthermore, we straightforwardly find that 
\begin{equation}
\begin{aligned}
&|R(t,\tau)|\\
=  &\frac{1}{T}\left|\int_{-\infty}^{\infty}\left(\begin{matrix}
 g(\zeta)g(\zeta-\tau)\\ \cdot
 (u(\zeta-t+\frac{T}{2})-u(\zeta-t-T-\frac{T}{2}))
\end{matrix} \right)d\zeta\right|\\
=&:|R(t,\tau)| _{max},
\end{aligned}
\label{eq: max abs gps autocor}
\end{equation}
where the equality is achieved at $\Delta f=0$.

According to Parseval principle, we have
\begin{equation}
\begin{aligned}
|R(t,\tau)| _{max} = \left|\int_{-\infty}^{\infty}G^2(f)sinc(fT)df\right|,
\end{aligned}
\label{eq: max abs gps autocor2}
\end{equation}
where $G(f)$ is the Fourier transform of $g(t)$.

In the sense of 3dB bandwidth ($B_{3dB}=\frac{0.6034}{T}$) for the \textit{sinc} function, the equation above can be written as
\begin{equation}
\begin{aligned}
|R(t,\tau)| _{max} \cong \frac{1}{2}\left|\int_{-B_{3dB}}^{B_{3dB}}G^2(f)df\right|,
\end{aligned}
\label{eq: max abs gps autocor3}
\end{equation}

In real cases, $\Delta f$ can not exactly be equal to zero, but very small value, $\Delta f\ll \frac{B_{3dB}}{100}$. Then, we have
\begin{equation}
\begin{aligned}
|R(t,\tau)| _{max} = &\frac{1}{2}\left|\int_{-\infty}^{\infty}G^2(f)\delta(f-\Delta f)sinc(fT)df\right|\\
\cong&\left|\int_{-B_{3dB}}^{B_{3dB}}G^2(f-\Delta f)df\right|.
\end{aligned}
\label{eq: max abs gps autocor4}
\end{equation}

From (\ref{eq: max abs gps autocor4}), we can adjust the value of $f_{dl}$ to make it approach to $f_d$ such that $\Delta f$ becomes small. Since the small $\Delta$ induces the maximum absolute correlation value, the Doppler shift can be determined by searching the maximum value of $|R(t,\tau)| _{max}$.

\subsection{Doppler Estimation Based on Cyclic-spectrum}

In this subsection, we introduce a method of estimating Doppler based on the cyclostatistics of GPS signal. The cyclostatistic of the received signal at cyclic frequency $\alpha$ is calculated as follows,
\begin{equation}
\begin{aligned}
&R^{\alpha}_R(\tau) = \lim_{T\rightarrow\infty}\frac{1}{T}\int_{-\frac{T}{2}}^{\frac{T}{2}}r(\zeta+\frac{\tau}{2})r(\zeta-\frac{\tau}{2})e^{-j2\pi\alpha\zeta}d\zeta\\
&= \lim_{T\rightarrow\infty}\frac{1}{T}\int_{-\frac{T}{2}}^{\frac{T}{2}}\begin{matrix}
\left(\begin{matrix}
\left(g(\zeta+\frac{\tau}{2})e^{-j2\pi f_d(\zeta+\frac{\tau}{2})}+\xi(\zeta+\frac{\tau}{2})\right)\\ \cdot\left(g(\zeta-\frac{\tau}{2})e^{-j2\pi f_d(\zeta-\frac{\tau}{2})}+\xi(\zeta-\frac{\tau}{2})\right)
\end{matrix} \right)\\ \cdot e^{-j2\pi\alpha\zeta}d\zeta
\end{matrix} \\
&\overset{(a)}{=}\lim_{T\rightarrow\infty}\frac{1}{T}\int_{-\frac{T}{2}}^{\frac{T}{2}}g(\zeta+\frac{\tau}{2})g(\zeta-\frac{\tau}{2})e^{-j2\pi(\alpha-2f_d)\zeta}d\zeta.\\
\end{aligned}
\label{eq: R cyclo alpha}
\end{equation}
where $(a)$ follows that the cyclostatistics of the noise at high order cyclic frequency is equal to zero. 

With the similar derivations, we calculate $R^{\alpha, G}_R(\tau)$ as follows,
\begin{equation}
\begin{aligned}
R^{\alpha}_{R,G}(\tau) =\lim_{T\rightarrow\infty}\frac{1}{T}\int_{-\frac{T}{2}}^{\frac{T}{2}}\left(\begin{matrix}
(g(\zeta+\frac{\tau}{2})g(\zeta-\frac{\tau}{2}))\\ \cdot e^{-j2\pi(\alpha-f_d-f_{dl})\zeta}
\end{matrix} \right) d\zeta.\\
\end{aligned}
\label{eq: RG cyclo alpha}
\end{equation}

Afterwards, we calculate the correlation between the cyclostatistics $R^{\alpha}_{R,G}$ and $R^{\alpha}_{R}$ which is taken at the statistics used for Doppler estimation,
\begin{equation}
\begin{aligned}
D_{ec}=\frac{1}{\Delta t} =\int_{\Delta t}R^{\alpha}_{R,G}(\tau){R^{\alpha}_{R}(\tau)}^*d\tau.
\end{aligned}
\label{eq: cyclo alpha corr}
\end{equation}

From (\ref{eq: cyclo alpha corr}), the absolute of $D_{ec}$ reaches the maximum at $f_d = f_{dl}$. Based on the results, we are able to calculate $D_{ec}$ at different $f_{dl}$'s and select the largest $|D_{ec}|$. The corresponding $f_{dl}$ is the estimation of $f_d$, that is,
\begin{equation}
\begin{aligned}
\hat{f}_d=\arg\max_{f_{dl}}\{D_{ec}(f_{dl})\}.
\end{aligned}
\label{eq: fd est cyclo}
\end{equation}

\section{Joint Estimation of Initial PN Code Phase and Doppler Based on Cyclic-spectrum}

In this section, the cyclic-spectrum of GPS signal is utilized to estimate the initial PN code phase and the Doppler shift in a joint way. Let $l(t)$ denote the locally generated GPS signal with a combination of configurable initial phase and Doppler shift. Let $R^{\alpha}_{R,L}(\tau)$ denote the cyclostatistics  calculated from $r(t)$ and $l(t)$ at the cyclic frequency of $\alpha$. $R^{\alpha}_{R,L}(\tau)$ is calculated as follows,
\begin{equation}
\begin{aligned}
&R^{\alpha}_{R,L}(\tau) = \lim_{T\rightarrow\infty}\frac{1}{T}\int_{-\frac{T}{2}}^{\frac{T}{2}}r(\zeta+\frac{\tau}{2})l(\zeta-\frac{\tau}{2})e^{-j2\pi\alpha\zeta}d\zeta\\
&= \lim_{T\rightarrow\infty}\frac{1}{T}\int_{-\frac{T}{2}}^{\frac{T}{2}}\left[g(\zeta+\frac{\tau}{2})+\xi(\zeta+\frac{\tau}{2})\right]l(\zeta-\frac{\tau}{2})e^{-j2\pi\alpha\zeta}d\zeta\\
&\overset{(a)}{=} \lim_{T\rightarrow\infty}\frac{1}{T}\int_{-\frac{T}{2}}^{\frac{T}{2}}g(\zeta+\frac{\tau}{2})l(\zeta-\frac{\tau}{2})e^{-j2\pi\alpha\zeta}d\zeta\\
&= \lim_{T\rightarrow\infty}\frac{1}{T}\int_{-\frac{T}{2}}^{\frac{T}{2}}\left(\begin{matrix}
(g(\zeta+\frac{\tau}{2}-D)e^{j2\pi\alpha(\zeta+\frac{\tau}{2})})\\ \cdot(g(\zeta-\frac{\tau}{2}-\Theta)e^{j2\pi\alpha(\zeta-\frac{\tau}{2})})
\end{matrix} \right) e^{-j2\pi\alpha\zeta}d\zeta\\
&= \lim_{T\rightarrow\infty}\frac{1}{T}\int_{-\frac{T}{2}}^{\frac{T}{2}}\begin{matrix}
\left(\begin{matrix}
g(\zeta-\frac{D-\Theta}{2}+(\frac{\tau}{2}-\frac{D-\Theta}{2}))\\ 
\cdot g(\zeta-\frac{D-\Theta}{2}-(\frac{\tau}{2}-\frac{D-\Theta}{2}))
\end{matrix} \right)\\
 \cdot e^{j2\pi(f_d+f_{dl}-\alpha)\zeta}d(\zeta-\frac{D-\Theta}{2})
\end{matrix} \\
&=R_G^{f_d+f_{dl}-\alpha}(\tau-D-\Theta)e^{j2\pi(f_d+f_{dl}-\alpha)\frac{D-\Theta}{2}},
\end{aligned}
\label{eq: R,L cyclo alpha}
\end{equation}
where $(a)$ follows that the cyclostatistics of the noise at high order cyclic frequency is equal to zero. 

Similarly, the cyclostatistics $R_R^{\alpha}$ is calculated as follows,
\begin{equation}
\begin{aligned}
&R^{\alpha}_{R}(\tau) = \lim_{T\rightarrow\infty}\frac{1}{T}\int_{-\frac{T}{2}}^{\frac{T}{2}}r(\zeta+\frac{\tau}{2})r(\zeta-\frac{\tau}{2})e^{-j2\pi\alpha\zeta}d\zeta\\
&= \lim_{T\rightarrow\infty}\frac{1}{T}\int_{-\frac{T}{2}}^{\frac{T}{2}}\left(\begin{matrix}
(g(\zeta+\frac{\tau}{2}-D)e^{j2\pi\alpha(\zeta+\frac{\tau}{2})})\\
 \cdot (g(\zeta-\frac{\tau}{2}-\Theta)e^{j2\pi\alpha(\zeta-\frac{\tau}{2})})
\end{matrix} \right) e^{-j2\pi\alpha\zeta}d\zeta\\
&= \lim_{T\rightarrow\infty}\frac{1}{T}\int_{-\frac{T}{2}}^{\frac{T}{2}}\begin{matrix}
g(\zeta-D+\frac{\tau}{2})g(\zeta-D-\frac{\tau}{2})\\
\cdot e^{j2\pi(2f_d-\alpha)\zeta}d(\zeta-D)
\end{matrix}\\
&=R_G^{2f_d-\alpha}(\tau)e^{j2\pi(2f_d-\alpha)D}.
\end{aligned}
\label{eq: R cyclo alpha joint}
\end{equation}

The corresponding cyclic-spectrum is calculated as follows,
\begin{equation}
\begin{aligned}
&S^{\alpha}_{R,L}(f)=\mathcal{F}\{R^{\alpha}_{R,L}(\tau)\}\\
&=\mathcal{F}\{R^{f_d+f_{dl}-\alpha}_{G}(\tau-D-\Theta)e^{j2\pi(f_d+f_{dl}-\alpha)\frac{D-\Theta}{2}}\}\\
&=S^{f_d+f_{dl}-\alpha}_{G}(f)e^{-j2\pi f(D-\Theta)}e^{j2\pi(f_d+f_{dl}-\alpha)\frac{D-\Theta}{2}},
\end{aligned}
\label{eq: S RL alpha}
\end{equation}
and 
\begin{equation}
\begin{aligned}
&S^{\alpha}_{R,L}(f)=\mathcal{F}\{R^{\alpha}_{R}(\tau)\}\\
&=\mathcal{F}\{R^{f_d+f_{dl}-\alpha}_{G}(\tau)e^{j2\pi(2f_d-\alpha)D}\}\\
&=S^{2f_d-\alpha}_{G}(f)e^{j2\pi(2f_d-\alpha)D}.
\end{aligned}
\label{eq: S R alpha}
\end{equation}

With the calculated cyclic-spectrum $S^{\alpha}_{R,L}$ and $S^{\alpha}_{R}$, we calculate the statistics used for joint estimation as follows,
\begin{equation}
\begin{aligned}
&\Lambda_{de}=\int_{-\infty}^{\infty}S^{\alpha}_{R,L}(f){S^{\alpha}_{R}}^*(f)df\\
&=e^{j2\pi(f_d+f_{dl}-\alpha)\frac{D-\Theta}{2}}e^{j2\pi(2f_d-\alpha)D}\\
\cdot&\int_{-\infty}^{\infty}S^{f_d+f_{dl}-\alpha}_{G}(f){S^{2f_d-\alpha}_{G}(f)}^*e^{-j2\pi f(D-\Theta)}df\\
\end{aligned}
\label{eq: detection var}
\end{equation}

The absolute value of $\Lambda_{dec}$,
\begin{equation}
\begin{aligned}
|\Lambda_{de}|=\left|\int_{-\infty}^{\infty}S^{f_d+f_{dl}-\alpha}_{G}(f){S^{2f_d-\alpha}_{G}(f)}^*e^{-j2\pi f(D-\Theta)}df\right|,
\end{aligned}
\label{eq: detection var abs}
\end{equation}
is used for the peak searching. 

From (\ref{eq: detection var abs}), we can achieve the GPS acquisition by peak searching the $|\Lambda_{de}|$ over the two dimensions of initial phase and Doppler. More concretely, we granularly change is value of $\Theta$ and $f_{dl}$ and calculate the corresponding $|\Lambda_{de}(\Theta,f_{dl})|$. At the maximum value, the initial phase delay and Doppler shift of the GPS signal are obtained,
\begin{equation}
\begin{aligned}
(\hat{\Theta},\hat{f}_{dl})=\arg\max_{\Theta,f_{dl}}\{|\Lambda_{de}(\Theta,f_{dl})|\},
\end{aligned}
\label{eq: detection principle}
\end{equation}

\begin{figure}[h]
\vspace{0 mm}
 \centering
  \includegraphics[width=1\linewidth]{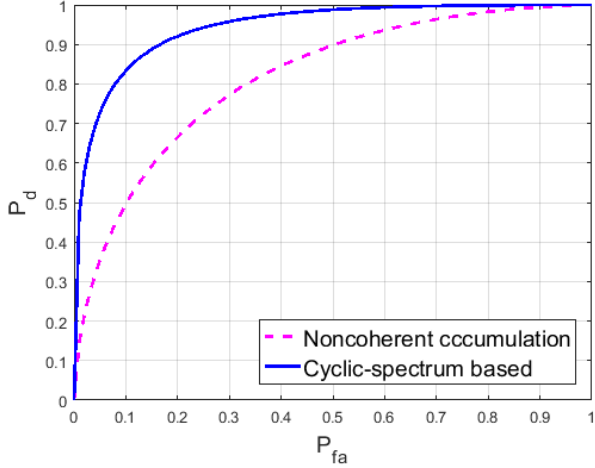}
 \caption{Receiver operating characteristic curve of the proposed cyclic-spectrum based GPS detection scheme}
\label{fig: roc curve}
\vspace{0 mm}
 \end{figure}

To test the cyclic-spectrum based GPS signal acquisition scheme, we first calculate the corresponding receiver operating characteristic curve~(ROC) which is plotted in Fig.~\ref{fig: roc curve}.

From Fig.~\ref{fig: roc curve}, the cyclic-spectrum based GPS acquisition scheme has better ROC curve which is more convex than the one for correlation based detection scheme. The advantage is due to the fact that GPS signal is immune to noise at high order cyclic frequency. 

\begin{figure}[h]
\vspace{0 mm}
 \centering
  \includegraphics[width=1\linewidth]{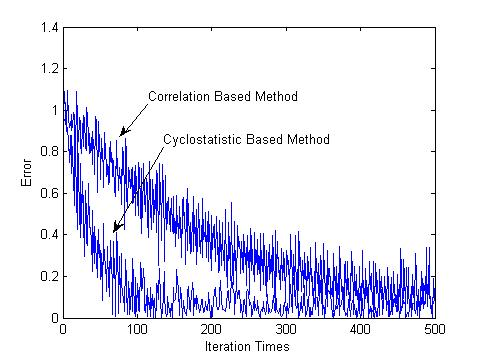}
  \caption{Error curve in iterative computations at 44dBHz}
\label{fig: err at 44}
\vspace{0 mm}
 \end{figure}

\begin{figure}[h]
\vspace{0 mm}
 \centering
  \includegraphics[width=1\linewidth]{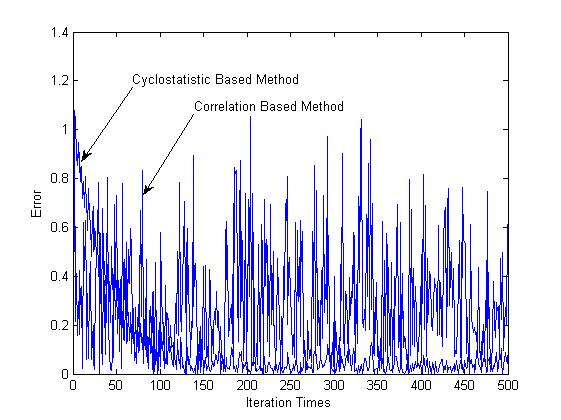}
 \caption{Error curve in iterative computations at 28dBHz}
\label{fig: err at 28}
\vspace{0 mm}
 \end{figure}

We also perform simulations to test the time delay estimation based on cyclic-spectrum of GPS signal which takes the form of iterative computation. Still, the correlation based data accumulation method is taken as the reference. In the experiment, the data length is equal to 20ms; iteration step is 0.03; sampling frequency is 5MHz. At the configuration, the corresponding value of $P$ is equals to 2500; the initial value of time delay $\hat{D}(0)$  is set to one.

Fig.~\ref{fig: err at 44} illustrates the error curve in the first 500 iteration times at the CNR of 44dBHz which corresponds to the LOS channel condition. From Fig.~\ref{fig: err at 44}, both the conventional method and the cyclostatistics based one can achieve successful estimation in high SNR condition, while the cyclostatistic based method converges faster than the conventional one. Fig.~\ref{fig: err at 28} shows the results at CNR 26dBHz which corresponds to the channel condition with severe degradation. In Fig.~\ref{fig: err at 28}, conventional method can not achieve the successful estimation on initial PN code phase while cyclostatistics based estimation method can.

\begin{figure}[h]
\vspace{0 mm}
 \centering
  \includegraphics[width=1\linewidth]{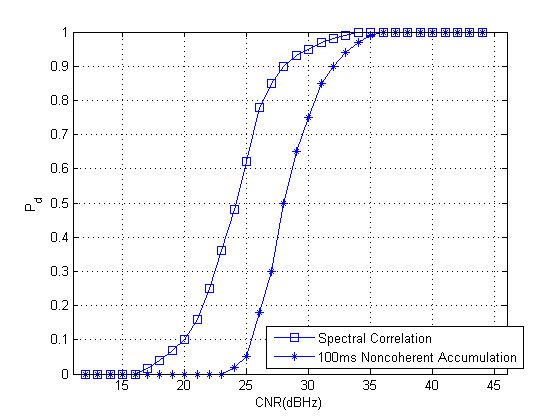}
 \caption{GPS signal detection performance comparison}
\label{fig: det prob}
\vspace{0 mm}
 \end{figure}

Furthermore, to test the performance at more signal conditions, detection probabilities at different CNR are calculated and the results are plotted in Fig.~\ref{fig: det prob}. As a contrast, the noncoherent detection is also performed. In the experiments the threshold is set aiming at one false alarm among $10^6$ times experiments which guarantees the false alarm probability at $10^{-6}$.

From Fig.~\ref{fig: det prob}, the cyclic-spectrum based method outperforms the noncoherent accumulation based method. The detection probability of the cyclic-spectrum based method approaches 0.9 at 28dBHz CNR, while the conventional method based on the 100ms noncoherent accumulation approaches 0.9 at 32dB/Hz. 

\section{Conclusions}

Since the mean of the GPS signal approaches zero and the correlation function is periodic, GPS signal is cyclostationary. We thus utilize the cyclostationary feature detect GPS signal acquisition at low CNR. Due to the non-cyclostationarity of Gaussian noise and the unconsistent cyclostationary feature of the jamming signal, the GPS signal can be distinguished from the background noise and the interference more easily.

The cyclostationary feature of the GPS signal is illustrated and the cyclic-spectrum of the GPS signal presented firstly; 

We first introduce how to utilize the cyclic-spectrum to estimate initial phase of pseudorandom code and Doppler shift respectively. An iterative estimation scheme is also proposed. Afterwards, a joint estimation scheme is presented. The simulation results show that the cyclic-spectrum based method outperforms the conventional structure which is based on noncoherent accumulation. 



{\vspace{-0mm}
\scriptsize
\bibliography{chann_emu_InfoCom2013}
}
\bibliographystyle{IEEEtran}

\end{document}